\documentclass[12pt,a4paper]{article}
\usepackage{amsmath,amsfonts,amsthm,graphicx}
\allowdisplaybreaks[3]
\numberwithin{equation}{section}

\usepackage[body={15.5cm,22.4cm}]{geometry}
\newcommand{\cc}{{\mathbf c}}
\newcommand{\cd}{{\mathbf c}^{\dagger}}
\newcommand{\n}{{\mathbf n}}

\newcommand{\Lf}{{\mathbf L}}
\newcommand{\Lbf}{\overline{\mathbf L}}
\newcommand{\Ls}{{\mathsf L}}
\newcommand{\Lbs}{\overline{\mathsf L}}

\newcommand{\Lfb}{\overline{\mathbf L}}
\newcommand{\Kf}{{\mathbf K}}
\newcommand{\Ks}{{\mathsf K}}
\newcommand{\eb}{\overline{e}}
\newcommand{\ez}{\mathfrak{e}}
\newcommand{\fz}{\mathfrak{f}}
\newcommand{\kz}{\mathfrak{k}}
\begin{document}
\title{
On diagonal solutions of the reflection equation
}
\author{Zengo Tsuboi
\\
Osaka city university advanced mathematical institute
\\
3-3-138 Sugimoto, Sumiyoshi-ku Osaka 558-8585, Japan
}
\maketitle
\begin{abstract}
We study solutions of the reflection equation 
associated with the quantum affine algebra 
$U_{q}(\hat{gl}(N))$ and obtain 
 diagonal K-operators in terms of the Cartan elements of 
 a quotient of $U_{q}(gl(N))$. 
We also consider intertwining relations for these  
K-operators and find an 
 augmented q-Onsager algebra like symmetry behind them. 
\end{abstract}
\vspace{6pt}

\noindent
Journal ref: 
Journal of Physics A: Mathematical and Theoretical 52 (2019) 155201
\vspace{6pt}

\noindent
DOI: https://dx.doi.org/10.1088/1751-8121/ab0b6d
\section{Introduction} 
The reflection equation \cite{Cherednik84} is a fundamental object in 
quantum integrable systems with open boundary conditions \cite{Skly}. 
It has the following form
\begin{align}
R_{12} \left(\frac{y}{x}\right) K_{1}(x) R_{21} \left( xy \right) 
 K_{2}(y) 
=K_{2}(y) 
 R_{12} \left(\frac{1}{xy} \right)  K_{1}(x) R_{21} \left(\frac{x}{y}\right), 
\quad  x,y \in \mathbb{C}.
\label{refeq00}
\end{align}
Here $R(x)$ is a solution (R-matrix) of the Yang-Baxter equation and $K(x)$ is 
 a K-matrix. The indices 1,2 denote the space where the operators act non-trivially. 
 In particular, 
there is a $N \times N$ diagonal matrix solution \cite{deVR93} of the reflection equation 
associated with the R-matrices \cite{Cherednik80} 
for the $N$-dimensional fundamental representation  of  $U_{q}(\hat{gl}(N))$. 
These R-matrices are specialization of more general operators 
called $L$-operators:  
$\Lf_{12}(x),\overline{\Lf}_{12}(x)  \in U_{q}(gl(N)) \otimes \mathrm{End}({\mathbb C}^{N})$. 
Namely, they are given by 
$R_{12}(x)=(\pi \otimes 1)\Lf_{12}(x)$, $R_{21}(x)=(\pi \otimes 1)\Lfb_{12}(x)$, 
 where $\pi$ is the fundamental representation of $U_{q}(gl(N))$.
In this context, 
a natural problem is to investigate the solutions of the reflection equation associated with 
the L-operators: 
\begin{align}
\Lf_{12} \left(\frac{y}{x}\right) \Kf_{1}(x) \overline{\Lf}_{12} \left( xy \right) 
 K_{2}(y) 
=K_{2}(y) 
 \Lf_{12} \left(\frac{1}{xy} \right)  \Kf_{1}(x) \overline{\Lf}_{12} \left(\frac{x}{y}\right),
\label{refeq20}
\end{align}
where the K-operator $\Kf(x)$ is an operator  in $U_{q}(gl(N))$. 
In this paper, we propose diagonal solutions of \eqref{refeq20}, namely  solutions written 
in terms of only the Cartan elements of (a quotient of) $U_{q}(gl(N))$. 
Evaluation of the K-operator for 
the fundamental representation of $U_{q}(gl(N))$ reproduces the $N \times N$ 
diagonal K-matrix: $K(x)=\pi(\Kf(x))$. 
In the context of Baxter Q-operators for integrable systems with open boundaries, 
 the essentially same K-operator for $N=2$ case was previously proposed in \cite{PT18}. 
This paper extends this to $N \ge 3$ case in part. 
We remark that K-operators for Baxter Q-operators appeared first in \cite{FS15} 
for the XXX-model ($q=1$ and $N=2$ case). 

We also rewrite \eqref{refeq20} as intertwining relations for the K-operator 
and speculate an explicit form of the underlying symmetry algebra. For $N=2$ case, 
it is known that the augmented q-Onsager algebra \cite{IT,BB1}, which is a co-ideal subalgebra of 
$U_{q}(\hat{sl}(2))$, serves as this. 
We find an augmented q-Onsager algebra like symmetry still exists for 
an oscillator representation even for $N\ge 3$ case. 
At a little more abstract level, the relevant algebras will be the reflection equation algebras \cite{Skly} 
and some coideal subalgebras of quantum affine algebras \cite{MRS}. 
In addition, symmetries of the transfer matrix of open spin chains with diagonal K-matrices are    discussed in \cite{DN98}. Intertwining relations for K-matrices are studied, 
for example, in \cite{DM03}.
However, an explicit expression for the higher rank analogue of the augmented q-Onsager algebra 
seems to be missing in the literatures. 
We expect that our results can be a cue for further study on this. 

There are several versions of the reflection equation (`left', `right', `twisted', `untwisted'). 
However, their solutions, K-matrices, 
are considered to be related each other by some transformations. Then 
we concentrate on one of them. 
Throughout this paper, we use the general 
gradation\footnote{We emulate \cite{BGKNR10} and their subsequent papers for this.} 
of $U_{q}(\hat{gl}(N))$. 
This does not produce particularly new results since 
 the difference of the gradation reduces to 
 a simple similarity transformation and a rescaling of the spectral parameter 
 on the level of the R-matrices. However, there is a merit to use it since 
the parameters in the general gradation play a role as `markers' and 
make it easier to trace the actions of the generators of $U_{q}(\hat{gl}(N))$. 
\section{The quantum affine and finite algebras}
In this section, we review the quantum algebras \cite{Dr85,Jimbo85,Jimbo86} of type A 
and associated L-operators. We also refer the books \cite{CP95,KS97} for review of this subject. 
\subsection{The quantum affine algebra $U_{q}(\hat{gl}(N))$ } 
We introduce a q-commutator 
$[X,Y]_{q}=XY-q YX$, and set  
$[X,Y]_{1}=[X,Y]$. 
The quantum affine algebra $U_{q}(\hat{gl}(N))$ 
is a Hopf algebra 
generated by the generators
\footnote{In this paper, we do not use the degree operator $d$. 
We will only consider  level zero representations. 
The notation $e_{0},f_{0}$ conventionally used in literatures corresponds to $e_{N},f_{N}$. 
We assume that the deformation parameter $q=e^{\hbar}$  ($\hbar \in \mathbb{C}$) 
is not a root of unity.}
 $e_{i},f_{i},k_{i}$, where 
 $i \in \{1,\dots, N\}$.  
For $i,j \in \{1,2,\dots,N \}$, the
 defining relations
  of the algebra are 
given by 
\begin{align}
\begin{split}
& [k_{i} ,k_{j}]=0, \qquad [k_{i},e_{j}] =(\delta_{ij} -\delta_{i,j+1} )e_{j}, 
\qquad [k_{i},f_{j}] =-(\delta_{ij} -\delta_{i,j+1} )f_{j}, 
\\
&[e_{i},f_{j}]=\delta_{ij} \frac{q^{h_{i}} -q^{-h_{i}} }{q-q^{-1}}, 
\\
& [e_{i},e_{j}]= [f_{i},f_{j}]=0 \quad \text{for} \quad |i-j|\ge 2, 
\\[6pt]
&[e_{i},[e_{i},[e_{i},e_{j}]_{q^{2}}]]_{q^{-2}}=
[f_{i},[f_{i},[f_{i},f_{j}]_{q^{-2}}]]_{q^{2}}=0 
\quad \text{for} \quad N=2 ,
\quad i \ne j ,
\\[6pt]
&[e_{i},[e_{i},e_{j}]_{q}]_{q^{-1}}=
[f_{i},[f_{i},f_{j}]_{q^{-1}}]_{q}=0 
\quad \text{for} \quad N \ge 3 ,
\quad |i-j|=1 ,
\end{split}
\label{rel-afglN}
\end{align}
where $h_{i}= k_{i} -  k_{i+1}$, and 
 $i,j$ should be interpreted  modulo $N$: 
($N+1 \equiv 1$).
 $\hat{c} =\sum_{i=1}^{N} k_{i } $ is a central element of the algebra. 
The algebra has the
 co-product $ \Delta : U_{q}(\hat{gl}(N)) \to U_{q}(\hat{gl}(N)) \otimes U_{q}(\hat{gl}(N))$ defined by 
\begin{align}
\begin{split}
\Delta (e_{i})&=e_{i} \otimes 1 + q^{-h_{i}} \otimes e_{i}, \\[6pt]
\Delta (f_{i})&=f_{i} \otimes q^{h_{i}} + 1 \otimes f_{i}, \\[6pt]
\Delta(k_{i}) &=k_{i} \otimes 1 + 1 \otimes k_{i}. 
\end{split}
\end{align} 
We will also use the opposite co-product defined by
\begin{align}
\Delta'=\sigma\circ \Delta,\qquad \sigma\circ 
(X\otimes Y)=
Y\otimes X,\qquad X,Y\in U_{q}(\hat{gl}(N)).
\end{align}
In addition to these, there are anti-poide and co-unit, which will not be used here. 
The Borel subalgebras ${\mathcal B}_{+}$ 
(resp. ${\mathcal B}_{-}$) is generated by the elements 
$e_{i}, k_{i} $ (resp. $f_{i},k_{i}$), where 
$i \in \{1,\dots, N\}$.


There exists a unique element \cite{Dr85,KT92} 
${\mathcal R} \in {\mathcal B}_{+} \otimes {\mathcal B}_{-} $ 
called the universal R-matrix which satisfies the following 
relations
\begin{align}
\begin{split}
\Delta'(a)\ {\mathcal R}&={\mathcal R}\ \Delta(a)
\qquad \text{for} \quad \forall\ a\in U_{q}(\hat{gl}(N)),   \\[6pt]
(\Delta\otimes 1)\, 
{\mathcal R}&={\mathcal R}_{13}\, {\mathcal R}_{23}, \\[6pt]
(1\otimes \Delta)\, {\mathcal R}&={\mathcal R}_{13}
{\mathcal R}_{12} .
\end{split}
\label{R-def}
\end{align}
where
\footnote{We will use similar notation for the L-operators 
to indicate the space on which they non-trivially act.} 
${\cal R}_{12}={\cal R}\otimes 1$, ${\cal R}_{23}=1\otimes {\cal R}$,
${\cal R}_{13}=(\sigma\otimes 1)\, {\cal R}_{23}$.
The Yang-Baxter equation 
\begin{align}
{\mathcal R}_{12}{\mathcal R}_{13}{\cal R}_{23}=
{\mathcal R}_{23}{\mathcal R}_{13}{\mathcal R}_{12} \label{YBE}
\end{align}
is a corollary of these relations \eqref{R-def}. 
The universal R-matrix can be written in the form
\begin{align}
{\mathcal R}=\overline{{\mathcal R}}\ q^{\sum_{i=1}^{N} k_{i} \otimes k_{i}}.
\label{R-red}
\end{align}
Here $\overline{{\mathcal R}}$  is the reduced universal $R$-matrix, which  
is a series in $e_j\otimes 1$ and $1 \otimes f_j$ 
and does not contain Cartan elements.  

\subsection{The quantum finite algebra $U_{q}(gl(N))$}
The finite quantum algebra $U_{q}(gl(N))$ is generated 
by the generators $\{ e_{i,i+1},e_{i+1,i} \}_{i=1}^{N-1}$ and $\{ e_{ii} \}_{i=1}^{N}$ 
obeying the following defining relations:
\begin{align}
& [e_{ii},e_{jj}]=0, \quad 
[e_{ii},e_{j,j+1}]= (\delta_{i,j}-\delta_{i,j+1} ) e_{j,j+1}, 
\quad 
[e_{ii},e_{j+1,j}]= -(\delta_{i,j}-\delta_{i,j+1} ) e_{j+1,j}, 
\nonumber
\\ & 
[e_{i,i+1}, e_{j+1,j}] = 
\delta_{ij} 
 \frac{q^{ e_{ii}- e_{i+1,i+1}} - 
q^{- e_{ii}+ e_{i+1,i+1}}}{q-q^{-1}} , 
\nonumber
\\
& [e_{i,i+1},e_{j,j+1}]=
[e_{i+1,i},e_{j+1,j}]=0 \quad \text{for} 
\quad |i-j|\ge 2, 
\\
& [e_{i,i+1},[e_{i,i+1},e_{j,j+1}]_{q}]_{q^{-1}}=
[e_{i+1,i},[e_{i+1,i},e_{j+1,j}]_{q^{-1}}]_{q}=0 
\quad \text{for}  \quad |i-j|= 1  .
\nonumber 
\end{align}
Note that $c=e_{11}+e_{22}+\cdots + e_{NN}$ is a central element of the algebra. 
We will also use elements defined recursively by 
\begin{align}
\begin{split}
e_{ij}&=[e_{ik},e_{kj}]_{q} \qquad 
\text{for} \quad i>k>j, \\
e_{ij}&=[e_{ik},e_{kj}]_{q^{-1}} \qquad 
\text{for} \quad i<k<j,
\end{split}
\label{eij}
\\[6pt]
\begin{split}
\overline{e}_{ij}&=[\overline{e}_{ik},\overline{e}_{kj}]_{q^{-1}} \qquad 
\text{for} \quad i>k>j, \\
\overline{e}_{ij}&=[\overline{e}_{ik},\overline{e}_{kj}]_{q} \qquad 
\text{for} \quad i<k<j,
\end{split}
\label{eijb}
\end{align}
where $\overline{e}_{i,i+1}=e_{i,i+1}$, $\overline{e}_{i+1,i}=e_{i+1,i}$.
%
We summarize relations among these elements in 
 the appendix. 

There is an evaluation map 
 $\mathsf{ev}_{x}$: 
$U_{q}(\hat{gl}(N)) \mapsto U_{q}(gl(N))$:  
\begin{align}
\begin{split}
& e_{N} \mapsto x^{s_{N}} q^{-e_{11}} e_{N1} q^{-e_{NN}},   \\
& f_{N} \mapsto x^{-s_{N}} q^{e_{NN}} e_{1N} q^{e_{11}}, 
 \\
& e_{i} \mapsto x^{s_{i}}e_{i,i+1}, \qquad 
f_{i} \mapsto x^{-s_{i}}e_{i+1,i} 
   \qquad \text{for} \quad 1 \le i \le N-1,
 \\
& 
k_{i} \mapsto e_{ii} \qquad \text{for} \quad 1 \le i \le N, 
\end{split} 
\label{eva}
\end{align}
where $x  \in {\mathbb C}$ is a spectral parameter; and the parameters 
$s_{i} \in {\mathbb Z}$ fix the gradation of the algebra. 
We will also use another 
 evaluation map 
 $\overline{\mathsf{ev}}_{x}$: 
$U_{q}(\hat{gl}(N)) \mapsto U_{q}(gl(N))$, 
which is defined by 
\begin{align}
\begin{split}
& e_{N} \mapsto x^{s_{N}} q^{e_{11}} \overline{e}_{N1} q^{e_{NN}},   \\
& f_{N} \mapsto x^{-s_{N}} q^{-e_{NN}} \overline{e}_{1N} q^{-e_{11}}, 
 \\
& e_{i} \mapsto x^{s_{i}}e_{i,i+1}, \qquad 
f_{i} \mapsto x^{-s_{i}}e_{i+1,i} 
   \qquad \text{for} \quad 1 \le i \le N-1,
 \\
& 
k_{i} \mapsto e_{ii} \qquad \text{for} \quad 1 \le i \le N.
\end{split} 
\label{eva2}
\end{align}
 
\subsection{Representations of quantum algebras}
Let $\pi$ be the fundamental representation of $U_{q}(gl(N))$, and 
$E_{ij}$ be a $N \times N$ matrix unit whose 
$(k,l)$-element is $\delta_{i,k} \delta_{j,l}$. We chose the basis of the representation 
space so that   
$\pi(e_{ij})=\pi(\eb_{ij})=E_{ij}$ holds. 
The composition $\pi_{x}=\pi \circ \mathsf{ev}_{x}=\pi \circ \overline{\mathsf{ev}}_{x}$ 
gives an evaluation representation
\footnote{Evaluation representations based the map $\mathsf{ev}_{x}$ 
 do not necessary coincide with the ones based on the map 
  $\overline{\mathsf{ev}}_{x}$ for more general representations.} of $U_{q}(\hat{gl}(N))$. 

The $q$-oscillator algebra is generated by the 
generators $\cc_{i},\cd_{i}, \n_{i}$ 
($i \in \{1,2,\dots, N\} $)  
 obeying the defining relations: 
\begin{align}
\begin{split}
& [\cc_{i}, \cd_{j}]_{q^{\delta_{ij}}}  =\delta_{ij} q^{-  \n_{i}}, 
\quad 
 [\cc_{i}, \cd_{j}]_{q^{-\delta_{ij}}}  =\delta_{ij} q^{\n_{i}} , 
\\
&  
[\n_{i}, \cc_{j}]=-\delta_{ij}\cc_{j}, \quad 
 [\n_{i}, \cd_{j}]=\delta_{ij}\cd_{j}, \quad
[\n_{i}, \n_{j}]=[\cc_{i}, \cc_{j}]=[\cd_{i}, \cd_{j}]=0.
\end{split}
\label{qosc}
\end{align}
A Fock space is given by the actions of the generators on the vacuum vector 
defined by $\cc_{i} |0  \rangle =\n_{i} |0  \rangle =0$. 
A highest weight representation of  $U_{q}(gl(N))$ with the highest weight 
$(m,0,\dots, 0)$ and the highest weight vector $ |0  \rangle$ 
($ e_{ii} |0 \rangle =m\delta_{i,1} |0 \rangle,  $
$e_{j,j+1} |0 \rangle =0 $, 
 $1 \le i \le N$, $1 \le j \le N-1$) 
 can be realized in terms of the q-oscillator algebra 
 (cf. q-analogue of the Holstein-Primakoff realization \cite{Palev98,Hayashi90}) : 
\begin{align}
\begin{split}
& e_{11}=m -\n_{2}-\dots - \n_{N}, 
\\
& e_{kk}=\n_{k},
\qquad \text{for} \quad 2 \le k \le N
\\
& e_{1j}= \cc_{j} q^{\sum_{k=2}^{j-1} \n_{k}} ,
\quad 
\eb_{1j}= \cc_{j} q^{-\sum_{k=2}^{j-1} \n_{k}} 
\qquad \text{for} \quad 2 \le j \le N, 
\\ 
& e_{i1}=\cd_{i} 
\left[m-\sum_{k=2}^{N} \n_{k}\right]_{q}
q^{-\sum_{k=2}^{i-1} \n_{k} }  ,
\quad 
 \eb_{i1}=\cd_{i} 
\left[m-\sum_{k=2}^{N} \n_{k}\right]_{q}
q^{\sum_{k=2}^{i-1} \n_{k} }
\quad \text{for} \quad 2 \le i \le N, 
\\
& e_{ij}=
 \cd_{i} \cc_{j} 
q^{\sum_{k=i+1}^{j-1} \n_{k}},
\quad 
\eb_{ij}=
 \cd_{i} \cc_{j} 
q^{-\sum_{k=i+1}^{j-1} \n_{k}}
\qquad \text{for} \quad 2 \le i<j \le N, 
\\ 
& e_{ij}=\cd_{i} \cc_{j} 
q^{-\sum_{k=j+1}^{i-1} \n_{k}} ,
\quad
 \eb_{ij}=\cd_{i} \cc_{j} 
q^{\sum_{k=j+1}^{i-1} \n_{k}} 
\qquad \text{for} \quad 2 \le j<i \le N 
,
\end{split} 
\label{HP-rep}
\end{align}
where $[x]_{q}=(q^{x}-q^{-x})/(q-q^{-1})$. 
This representation is infinite dimensional and has an invariant 
subspace if $m$ is a non-negative integer. Factoring out the 
invariant subspace, one can obtain 
 the $m$-th symmetric tensor representation. 
 The K-operators in section 3 are valid at least for this 
 q-oscillator realization. 
 We also use this q-oscillator realization to check the 
 commutation relations for underlying symmetry in section 4. 
%
%
\subsection{L-operators}

The so-called L-operators are images of the universal R-matrix, which are given by 
$\Lf (xy^{-1})=\phi(xy^{-1})(\mathsf{ev}_{x} \otimes \pi_{y}) {\mathcal R}$, 
$\Lfb (xy^{-1})=\overline{\phi}(xy^{-1})(\overline{\mathsf{ev}}_{x} \otimes \pi_{y}) {\mathcal R}_{21}$, 
where $\phi(xy^{-1})$ and $\overline{\phi}(xy^{-1})$ are overall factors. 
They are solutions of the intertwining relations 
following from \eqref{R-def}:
\begin{align}
& \left((\mathsf{ev}_{x} \otimes \pi_{y})  \Delta^{\prime }(a)\right) \Lf (xy^{-1}) =\Lf (xy^{-1}) 
\left((\mathsf{ev}_{x} \otimes \pi_{y})  \Delta(a)\right)  ,
 \label{intert1}
\\[6pt]
& \left((\overline{\mathsf{ev}}_{x} \otimes \pi_{y})  \Delta(a)\right) \Lfb (xy^{-1}) =\Lfb (xy^{-1}) 
\left((\overline{\mathsf{ev}}_{x} \otimes \pi_{y})  \Delta^{\prime }(a)\right)  
\quad \forall a \in U_{q}(\hat{gl}(N)).
 \label{intert2}
\end{align}
One can solve
\footnote{Taking note on the expression \eqref{R-red}, 
we adopt the solutions which satisfy 
$\Lf(0)=\Lfb(\infty)=\sum_{i=1}^{N} q^{e_{ii}}\otimes E_{ii}$ 
for the case $s_{k}>0$ (for all $ k \in \{1,2,\dots, N\}$).}
 these to get 
(cf.\ \cite{Jimbo86,KP08})
\begin{align}
\Lf (xy^{-1}) &=\sum_{j,k=1}^{N}
\left\{ \left(xy^{-1}\right)^{-\xi_{k}+\xi_{j}} L^{+}_{kj} -
\left(xy^{-1}\right)^{s-\xi_{k}+\xi_{j}} L^{-}_{kj} 
\right\} \otimes E_{kj}, 
 \label{Lop}
\\[6pt]
\Lfb(xy^{-1}) &=\sum_{j,k=1}^{N}
\left\{- \left(xy^{-1}\right)^{-s-\xi_{k}+\xi_{j}}  \overline{L}^{+}_{kj} +
\left(xy^{-1}\right)^{-\xi_{k}+\xi_{j}} \overline{L}^{-}_{kj} 
\right\} \otimes E_{kj}, 
\label{Lbop}
\end{align}
where  $\xi_{k}=s_{k}+s_{k+1}+\cdots +s_{N}$, $s=\xi_{1}$, and 
the coefficients are related to the generators of  $U_{q}(gl(N))$  
as 
\begin{align}
\begin{split}
& L^{+}_{ii}=q^{e_{ii}}, 
\qquad 
L^{-}_{ii}=q^{-e_{ii}}, 
\\[6pt]
& L^{+}_{ij}=(q-q^{-1})e_{ji} q^{ e_{jj}} ,
\qquad  L^{-}_{ij}=0
\quad \text{for} \quad i >j, \\[6pt]
& L^{-}_{ij}=-(q-q^{-1})q^{- e_{ii}}e_{ji}  ,
\qquad  L^{+}_{ij}=0
\quad \text{for} \quad i <j ,
\end{split}
\label{maprll3}
\\[10pt]
\begin{split}
&\overline{L}^{+}_{ii}=q^{-e_{ii}}, 
\qquad 
\overline{L}^{-}_{ii}=q^{e_{ii}}, 
\\[6pt]
& \overline{L}^{+}_{ij}=-(q-q^{-1})q^{ -e_{ii}} \overline{e}_{ji} ,
\qquad  \overline{L}^{-}_{ij}=0
\quad \text{for} \quad i >j, \\[6pt]
& \overline{L}^{-}_{ij}=(q-q^{-1})\overline{e}_{ji}  q^{ e_{jj}},
\qquad  \overline{L}^{+}_{ij}=0 
\quad \text{for} \quad i <j .
\end{split}
\label{maprll3b}
\end{align}
We remark that 
the second L-operator \eqref{Lbop} follows from the first one  \eqref{Lop} by the 
transformation 
\begin{align}
\begin{split}
& e_{ij} \mapsto \eb_{N+1-i,N+1-j} , 
\quad 
E_{ij} \mapsto E_{N+1-i,N+1-j} 
\quad \text{for} \quad 1 \le i,j \le N,
\\[6pt] 
\quad 
&s_{l} \mapsto -s_{N-l} \quad \text{for} \quad 
\quad 1 \le l \le N-1, 
\\[6pt] 
\quad 
&
s_{N} \mapsto -s_{N},
\end{split}
\label{transLb}
\end{align}
which is composition of automorphisms of $U_{q}(\hat{gl}(N))$ and $U_{q}(gl(N))$.
In fact, the intertwining relation \eqref{intert2} follows from \eqref{intert1} 
by the same transformation \eqref{transLb}. 


Evaluating the L-operators for the fundamental representation
\footnote{Up to an overall factor, \eqref{Rb}
coincides with $(\pi_{x} \otimes \pi_{1}) {\mathcal R}_{21}$ 
since \eqref{ev-evb} and $c=1$ hold true for the fundamental representation.}, 
we obtain R-matrices  \cite{Cherednik80}, 
\begin{multline}
 R(x)=(\pi \otimes 1) \Lf(x)
= \sum_{i=1}^{N}\left( q - x^{s} q^{-1}\right)E_{ii} \otimes E_{ii} +
\sum_{i \ne j} \left( 1 - x^{s}\right) E_{ii} \otimes E_{jj} 
\\+
(q-q^{-1}) 
\sum_{i<j}  x^{\xi_{i}-\xi_{j}} E_{ij} \otimes E_{ji},
+
(q-q^{-1}) 
\sum_{i>j} x^{s+\xi_{i}-\xi_{j}} E_{ij} \otimes E_{ji} ,
 \label{RR}
\end{multline}
\begin{multline}
 \overline{R}(x)=(\pi \otimes 1) \Lfb(x)=
 \sum_{i=1}^{N}\left( q - x^{-s}q^{-1}\right)E_{ii} \otimes E_{ii} +
\sum_{i \ne j} \left( 1 - x^{-s}\right) E_{ii} \otimes E_{jj} 
\\+
(q-q^{-1}) 
\sum_{i>j}  x^{\xi_{i}-\xi_{j}} E_{ij} \otimes E_{ji},
+
(q-q^{-1}) 
\sum_{i<j} x^{-s+\xi_{i}-\xi_{j}} E_{ij} \otimes E_{ji} .
\label{Rb}
\end{multline}
%
\section{The reflection equation and its solutions}
\label{sec:RE}
In this section, we will derive intertwining relations  from the reflection equation 
associated with the L-operators, 
and obtain  K-operators in terms of the Cartan elements of 
(a quotient of) $U_{q}(gl(N))$. 

We start from the following form of the reflection equation for the R-matrices 
\eqref{RR} and \eqref{Rb}:
\begin{align}
R_{12} \left(\frac{y}{x}\right) K_{1}(x) \overline{R}_{12} \left( xy \right) 
 K_{2}(y) 
=K_{2}(y) 
 R_{12} \left(\frac{1}{xy} \right)  K_{1}(x) \overline{R}_{12} \left(\frac{x}{y}\right).
\label{refeq0}
\end{align}
It is known that 
this equation allows a $N \times N$ diagonal matrix solution (K-matrix) \cite{deVR93}. 
In our convention, it reads
\begin{align}
K(x)=\sum_{k=1}^{a}x^{2(s-\xi_{k})}(\epsilon_{-}+\epsilon_{+} x^s)E_{kk}+
\sum_{k=a+1}^{N}x^{2(s-\xi_{k})}(\epsilon_{-}+\epsilon_{+} x^{-s})E_{kk},
\label{K-dia}
\end{align}
where $a \in \{0,1,\dots, N \}$, $\epsilon_{\pm} \in {\mathbb C}$. 
Now we would like to consider the 
reflection equation for the L-operators \eqref{Lop} and \eqref{Lbop}: 
\begin{align}
\Lf_{12} \left(\frac{y}{x}\right) \Kf_{1}(x) \overline{\Lf}_{12} \left( xy \right) 
 K_{2}(y) 
=K_{2}(y) 
 \Lf_{12} \left(\frac{1}{xy} \right)  \Kf_{1}(x) \overline{\Lf}_{12} \left(\frac{x}{y}\right).
\label{refeq2}
\end{align}
The reflection equation \eqref{refeq0} is the image of  \eqref{refeq2} 
for $\pi \otimes 1$. 
Expanding \eqref{refeq2} with respect to $y$, we obtain 
\begin{align}
&\sum_{k=1}^{N}x^{-2\xi_{k}}
\left(
L^{+}_{ik} \Kf(x) \overline{L}^{+}_{kj}-L^{-}_{ik} \Kf(x) \overline{L}^{-}_{kj} 
\right)=0 
\quad \text{for} \quad i,j \le a 
\quad \text{or} \quad i,j >a, 
\label{REex1}
\\[6pt]
&\sum_{k=1}^{N}x^{-2\xi_{k}}
\left\{
\left(
x^{s}L^{+}_{ik} \Kf(x) \overline{L}^{-}_{kj}+x^{-s} L^{-}_{ik} \Kf(x) \overline{L}^{+}_{kj} 
\right)\epsilon_{+}+
 L^{-}_{ik} \Kf(x) \overline{L}^{-}_{kj} 
\epsilon_{-} 
\right\}=0 
\nonumber 
\\
&
\hspace{230pt} \text{for} \quad i \le a <j, 
 \label{REex2}
\\[6pt]
&\sum_{k=1}^{N}x^{-2\xi_{k}}
\left\{
\left(
x^{s}L^{+}_{ik} \Kf(x) \overline{L}^{-}_{kj}+x^{-s} L^{-}_{ik} \Kf(x) \overline{L}^{+}_{kj} 
\right)\epsilon_{+}+
 L^{+}_{ik} \Kf(x) \overline{L}^{+}_{kj} 
\epsilon_{-} 
\right\}=0 
\nonumber 
\\
&
\hspace{230pt}  \text{for} \quad j \le a <i . 
\label{REex3}
\end{align}
One can rewrite
\footnote{The commutation relations for the Cartan elements from \eqref{REex1} 
are of the form 
$q^{e_{ii}} \Kf(x) q^{-e_{ii}}= q^{-e_{ii}}\Kf(x) q^{e_{ii}}$, from which 
 $e_{ii} \Kf(x)= \Kf(x) e_{ii}$ are deduced 
(under expansion of $e_{ii}=\log q^{e_{ii}}/\log q$). 
We have used these to simplify \eqref{inter1}-\eqref{inter5}.}
 these in terms of $e_{ij}$ and $\eb_{ij}$: 
\begin{align}
& q^{2e_{ii}} \Kf(x)= \Kf(x) q^{2e_{ii}} ,
\label{rel-CartanK}
\\[6pt]
& x^{-2\xi_{j}}e_{ji} \Kf(x)-(q-q^{-1})\sum_{k=j+1}^{i-1}x^{-2\xi_{k}}e_{ki}\Kf(x)\eb_{jk}
-x^{-2\xi_{i}}\Kf(x) \eb_{ji} =0
\nonumber \\
& \hspace{180pt } \text{for} \quad j <i \le a \quad \text{or} \quad a <j<i, 
 \label{inter1}
\\[6pt]
& x^{-2\xi_{j}}e_{ji} \Kf(x)+(q-q^{-1})\sum_{k=i+1}^{j-1}x^{-2\xi_{k}}e_{ki}\Kf(x)\eb_{jk}
-x^{-2\xi_{i}}\Kf(x) \eb_{ji} =0
\nonumber \\
& \hspace{180pt } \text{for} \quad i <j \le a \quad \text{or} \quad a <i<j, 
 \label{inter2}
\\[6pt]
& x^{-2\xi_{j}}e_{ji}(\epsilon_{+}x^{-s}q^{-2e_{jj}}+\epsilon_{-}) \Kf(x)
 -
(q-q^{-1})\Bigl(
\epsilon_{+}x^{s}\sum_{k=1}^{i-1}x^{-2\xi_{k}}q^{e_{ii}}e_{ki}q^{e_{kk}}\Kf(x)
\eb_{jk}
\nonumber \\
&
\qquad  
+\epsilon_{+}x^{-s}\sum_{k=j+1}^{N}x^{-2\xi_{k}}e_{ki}\Kf(x)q^{-e_{kk}}\eb_{jk}q^{-e_{jj}}
-\epsilon_{-}\sum_{k=i+1}^{j-1}x^{-2\xi_{k}}e_{ki}\Kf(x)\eb_{jk}
\Bigr)
\nonumber \\
& \qquad  
-x^{-2\xi_{i}}\Kf(x) \eb_{ji}(\epsilon_{+}x^{s}q^{2e_{ii}-2}+\epsilon_{-}) =0
\quad  \text{for} \quad i \le a <j, 
\label{inter4}
\\[6pt]
& x^{-2\xi_{j}}e_{ji}(\epsilon_{+}x^{s}q^{2e_{jj}}+\epsilon_{-}) \Kf(x)
 +
(q-q^{-1})\Bigl(
\epsilon_{+}x^{s}\sum_{k=1}^{j-1}x^{-2\xi_{k}}e_{ki}q^{e_{kk}}\Kf(x)
\eb_{jk}q^{e_{jj}}
\nonumber \\
&
\qquad  
+\epsilon_{+}x^{-s}\sum_{k=i+1}^{N}x^{-2\xi_{k}}q^{-e_{ii}}e_{ki}\Kf(x)q^{-e_{kk}}\eb_{jk}
-\epsilon_{-}\sum_{k=j+1}^{i-1}x^{-2\xi_{k}}e_{ki}\Kf(x)\eb_{jk}
\Bigr)
\nonumber \\
& \qquad  
-x^{-2\xi_{i}}\Kf(x) \eb_{ji}(\epsilon_{+}x^{-s}q^{-2e_{ii}+2}+\epsilon_{-}) =0
\quad  \text{for} \quad j \le a <i. 
 \label{inter5}
\end{align}
For $i$ and $j$ satisfying $|i-j|=1$,  
the relations \eqref{inter1} and \eqref{inter2} reduce to 
\begin{multline}
  x^{-2\xi_{j}} e_{j,j+1} \Kf(x)  =x^{-2\xi_{j+1}}\Kf(x) e_{j,j+1} , 
  \quad 
    x^{-2\xi_{j+1}} e_{j+1,j} \Kf(x)  =x^{-2\xi_{j}}\Kf(x) e_{j+1,j} , 
\\
\text{for} \quad   j \ne a .
\label{rel-eK}
\end{multline} 
 Taking note on  \eqref{eij} and \eqref{eijb} (and \eqref{rel-CartanK}), one can derive
\begin{align}
& x^{-2\xi_{j}}\eb_{ji} \Kf(x)  =x^{-2\xi_{i}}\Kf(x) \eb_{ji} 
\quad  \text{for} \quad i,j \le a \quad \text{or} \quad a <i,j, 
 \label{inter1b}
\\[6pt]
& x^{-2\xi_{j}}e_{ji} \Kf(x)  =x^{-2\xi_{i}}\Kf(x) e_{ji} 
\quad  \text{for} \quad i,j \le a \quad \text{or} \quad a <i,j 
 \label{inter2b}
\end{align}
from \eqref{rel-eK}. 
Then the relations \eqref{inter1} and \eqref{inter2} boil down to 
the relations 
\begin{align}
& e_{ji}-(q-q^{-1})\sum_{k=j+1}^{i-1}e_{ki}\eb_{jk}-\eb_{ji} =0
\quad \text{for} \quad j<i,
\\[6pt]
& e_{ji}+(q-q^{-1})\sum_{k=i+1}^{j-1}e_{ki}\eb_{jk}-\eb_{ji} =0
\quad \text{for} \quad i<j ,
\end{align}
which are special cases of \eqref{eeb1} and \eqref{eeb1-2}, 
under \eqref{inter1b} and \eqref{inter2b}. 
Thus it suffices to consider \eqref{rel-eK} instead of  \eqref{inter1} and \eqref{inter2}.
We further rewrite \eqref{inter4} and \eqref{inter5}
 under the relations \eqref{inter1b} and \eqref{inter2b}, 
in the form of intertwining relations.
\\
For $i \le a <j$,  \eqref{inter4} reduces to 
\begin{align}
\begin{split}
&
x^{-2\xi_{j}}
(\epsilon_{+}x^{-s}A_{ji}+\epsilon_{-}B_{ji}) \Kf(x)=
x^{-2\xi_{i}}
\Kf(x)(\epsilon_{+}x^{s}C_{ji}+\epsilon_{-}D_{ji}), 
\\[6pt]
&\qquad A_{ji}=
e_{ji}q^{-2e_{jj}}
 -
(q-q^{-1})\sum_{k=j+1}^{N} e_{ki}\eb_{jk} q^{-e_{kk}-e_{jj}+1} , 
\\
&\qquad  B_{ji}=
e_{ji}+(q-q^{-1}) \sum_{k=a+1}^{j-1}e_{ki}\eb_{jk},
\\
&\qquad 
C_{ji}=
\eb_{ji}q^{2e_{ii}-2}
+(q-q^{-1}) \sum_{k=1}^{i-1} e_{ki}\eb_{jk} q^{e_{kk}+e_{ii}-2},
 \\
& \qquad  D_{ji}=\eb_{ji}
-(q-q^{-1}) \sum_{k=i+1}^{a} e_{ki}\eb_{jk},
\end{split}
\label{inter4a}
\end{align}
and for $j \le a <i$, \eqref{inter5} reduces to 
\begin{align}
\begin{split}
&
x^{-2\xi_{j}}
(\epsilon_{+}x^{s}A_{ji}+\epsilon_{-}B_{ji}) \Kf(x)=
x^{-2\xi_{i}}
\Kf(x)(\epsilon_{+}x^{-s}C_{ji}+\epsilon_{-}D_{ji}), 
\\[6pt]
&\qquad  A_{ji}=
e_{ji}q^{2e_{jj}}
 +
(q-q^{-1})\sum_{k=1}^{j-1} e_{ki}\eb_{jk} q^{e_{kk}+e_{jj}-1},
\\
&\qquad 
B_{ji}=e_{ji}
-(q-q^{-1})\sum_{k=j+1}^{a} e_{ki}\eb_{jk},
\\
&\qquad 
C_{ji}=\eb_{ji}q^{-2e_{ii}+2}
-(q-q^{-1}) \sum_{k=i+1}^{N} e_{ki}\eb_{jk} q^{-e_{kk}-e_{ii}+2},
\\
& \qquad 
D_{ji}=\eb_{ji}
+(q-q^{-1})\sum_{k=a+1}^{i-1} e_{ki}\eb_{jk} 
.
\end{split}
\label{inter4b}
\end{align}
We remark that the relation  $B_{ji}=D_{ji}$ holds 
due to \eqref{eeb1} and \eqref{eeb1-2}. 
%
We find that the following operators 
\footnote{In 2016, we were informed by S. Belliard that he found a (non-diagonal) solution for 
the rational case $Y(sl(2))$.}
 satisfy \eqref{rel-CartanK},  \eqref{rel-eK}, \eqref{inter4a} and \eqref{inter4b}  
for the q-oscillator representation \eqref{HP-rep}, and 
thus solve the reflection equation \eqref{refeq2}.
\begin{align}
\Kf(x)&=
q^{2(c-1)\sum_{k=1}^{a}e_{kk}}x^{2\sum_{k=1}^{N}(s\theta(1 \le k \le a)-\xi_{k})e_{kk}}
\frac{\left( -\frac{\epsilon_{-}}{\epsilon_{+}} 
x^{s} q^{2\sum_{k=a+1}^{N}e_{kk}-2};q^{-2} \right)_{\infty}}
{\left( -\frac{\epsilon_{-}}{\epsilon_{+}} 
x^{-s} q^{-2\sum_{k=1}^{a}e_{kk}};q^{-2} \right)_{\infty}}
\nonumber 
\\
& \hspace{244pt}  \text{for} \quad |q|>1, \quad \epsilon_{+}\ne 0, 
 \label{K-op}
 \\[6pt]
 \Kf(x)&=
q^{2(c-1)\sum_{k=1}^{a}e_{kk}}x^{2\sum_{k=1}^{N}(s\theta(1 \le k \le a)-\xi_{k})e_{kk}}
\frac{\left( -\frac{\epsilon_{-}}{\epsilon_{+}} 
x^{-s} q^{-2\sum_{k=1}^{a}e_{kk}+2};q^{2} \right)_{\infty}}
{\left( -\frac{\epsilon_{-}}{\epsilon_{+}} 
x^{s} q^{2\sum_{k=a+1}^{N}e_{kk}};q^{2} \right)_{\infty}}
\nonumber 
\\
& \hspace{244pt}  \text{for} \quad |q|<1,  \quad \epsilon_{+}\ne 0,
 \label{K-op2}
 \\[6pt]
\Kf(x)&=
x^{-2\sum_{k=1}^{N}\xi_{k}e_{kk}}
\frac{\left( -\frac{\epsilon_{+}}{\epsilon_{-}} 
x^{s} q^{2\sum_{k=1}^{a}e_{kk}-2};q^{-2} \right)_{\infty}}
{\left( -\frac{\epsilon_{+}}{\epsilon_{-}} 
x^{-s} q^{-2\sum_{k=a+1}^{N}e_{kk}};q^{-2} \right)_{\infty}}
\quad  \text{for} \quad |q|>1, \quad \epsilon_{-}\ne 0,
 \label{K-op3}
 \\[6pt]
\Kf(x) &=
x^{-2\sum_{k=1}^{N}\xi_{k}e_{kk}}
\frac{\left( -\frac{\epsilon_{+}}{\epsilon_{-}} 
x^{-s} q^{-2\sum_{k=a+1}^{N}e_{kk}+2};q^{2} \right)_{\infty}}
{\left( -\frac{\epsilon_{+}}{\epsilon_{-}} 
x^{s} q^{2\sum_{k=1}^{a}e_{kk}};q^{2} \right)_{\infty}}
\quad  \text{for} \quad |q|<1,  \quad \epsilon_{-}\ne 0,
 \label{K-op4}
\end{align}
where $c=\sum_{k=1}^{N}e_{kk}$, 
$\theta(\text{True})=1$, $\theta(\text{False})=0$, and 
$(x;q)_{\infty}=\prod_{j=0}^{\infty}(1-xq^{j})$. 
One can directly check that \eqref{K-op}-\eqref{K-op4} satisfy \eqref{rel-CartanK} and 
 \eqref{rel-eK} for the generic generators of $U_{q}(gl(N))$. 
 One can also show that 
the operators \eqref{K-op}-\eqref{K-op4} satisfy the following relations
\begin{align}
\Kf(x){\mathbf E}_{ji}&= x^{2(\xi_{i}-\xi_{j})}{\mathbf E}_{ji}\Kf(x)
\left(\frac{\epsilon_{+}x^{s}q^{2\sum_{k=1}^{a}e_{kk}}+\epsilon_{-}}{\epsilon_{+}x^{-s}q^{-2\sum_{k=a+1}^{N}e_{kk}+2}+\epsilon_{-}} \right) 
\quad \text{for} \quad j \le a < i,
\\[6pt]
\Kf(x){\mathbf E}_{ji}&= x^{2(\xi_{i}-\xi_{j})}{\mathbf E}_{ji}\Kf(x)
\left(\frac{\epsilon_{+}x^{-s}q^{-2\sum_{k=a+1}^{N}e_{kk}}+\epsilon_{-}}{\epsilon_{+}x^{s}q^{2\sum_{k=1}^{a}e_{kk}-2}+\epsilon_{-}} \right) 
\quad \text{for} \quad i \le a < j,
\end{align}
where ${\mathbf E}_{ji}$ is any operator obeying 
$q^{e_{kk}}{\mathbf E}_{ji}q^{-e_{kk}}=q^{\delta_{kj}-\delta_{ki}}{\mathbf E}_{ji}$. 
Thus \eqref{inter4a} and \eqref{inter4b} reduce to 
\begin{align}
B_{ji}&=A_{ji}q^{2\sum_{k=a+1}^{N}e_{kk}}=C_{ji}q^{-2\sum_{k=1}^{a}e_{kk}+2}
\quad \text{for} \quad i \le a < j,
\label{con1}
\\[6pt]
B_{ji}&=A_{ji}q^{-2\sum_{k=1}^{a}e_{kk}}=C_{ji}q^{2\sum_{k=a+1}^{N}e_{kk}-2}
\quad \text{for} \quad j \le a < i.
\label{con2}
\end{align}
These relations \eqref{con1} and \eqref{con2} 
give a constraint on the class of representations of $U_{q}(gl(N))$.  
In other words, we have to consider the quotient of $U_{q}(gl(N))$ 
by the relations \eqref{con1} and \eqref{con2}. 
Moreover, we find that the first evaluation map \eqref{eva} is 
almost
\footnote{This comes from  \eqref{con1} for $(i,j)=(1,N)$ and 
 \eqref{con2} for $(i,j)=(N,1)$ under $a\in \{1,2,\dots, N-1\}$. 
The factor $q^{\mp 2(c-1)}$ in the right hand side can be eliminated if one 
considers composition of the automorphism 
$e_{i}\mapsto q^{-2(\hat{c}-1)\delta_{iN}}e_{i}$, 
$f_{i}\mapsto q^{2(\hat{c}-1)\delta_{iN}}f_{i}$, 
$k_{i}\mapsto k_{i}$ 
 of $U_{q}(\hat{gl}(N))$ 
and the evaluation map $\overline{\mathsf{ev}}_{x}$.}
 equivalent to the second one \eqref{eva2} under these:
\begin{align}
\mathsf{ev}_{x}(e_{N})=q^{-2(c-1)}\overline{\mathsf{ev}}_{x}(e_{N}),
\quad
\mathsf{ev}_{x}(f_{N})=q^{2(c-1)}\overline{\mathsf{ev}}_{x}(f_{N})
\quad
\text{[under \eqref{con1} and \eqref{con2}]}.
 \label{ev-evb}
\end{align}
One can check that \eqref{HP-rep} satisfies \eqref{con1} and \eqref{con2}. 
In addition, the fundamental representation of $U_{q}(gl(N))$, which is a quotient of 
\eqref{HP-rep}  at $m=1$, also fulfills \eqref{con1} and \eqref{con2}. 
Then we recover \eqref{K-dia} 
from \eqref{K-op}:
\begin{align}
\pi(\Kf(x))=\frac{x^{-s}\left(-\frac{\epsilon_{-}}{\epsilon_{+}}x^{s}q^{-2} ;q^{-2}\right)_{\infty}}{\epsilon_{+}\left(-\frac{\epsilon_{-}}{\epsilon_{+}}x^{-s} ;q^{-2}\right)_{\infty}}K(x).
\end{align}
It is important to note that the relations  \eqref{con1} and \eqref{con2} 
become trivial for $N=2$ case. 
This means that the K-operators \eqref{K-op}-\eqref{K-op4} written in terms of the generic Cartan 
elements solve the reflection equation \eqref{refeq2} 
without any constraint on the representations 
for $U_{q}(gl(2))$ case. In fact, \eqref{K-op} for $N=2$ and $a=1$ reproduces 
\footnote{$x^{s_{2}c} q^{-\frac{(c-1)((s+s_{2})c+s_{1}H)}{s}}\Kf(xq^{-\frac{c-1}{s}})$ 
for $N=2$
coincides with eq.\ (4.9) in \cite{PT18}.} 
the K-operator proposed in \cite{PT18}. 

Let us consider a product of the L-operators
\begin{multline}
\Lf(x) q^{\frac{2(c-1)}{s}\sum_{j=1}^{N}\xi_{j}e_{jj}}
\Lfb(xq^{-\frac{2(c-1)}{s}})
q^{-\frac{2(c-1)}{s}\sum_{j=1}^{N}\xi_{j}e_{jj}}
=
\\[6pt]
=
\sum_{i=1}^{N}(-x^{s}-x^{-s}+G_{i})\otimes E_{ii} 
+
(q-q^{-1})\sum_{i<j}(C_{ji}-A_{ji}q^{2(c-1)})q^{e_{jj}-e_{ii}+1}\otimes E_{ij} 
\\[6pt]
+
(q-q^{-1})\sum_{i>j}(A_{ji}-C_{ji}q^{2(c-1)})\otimes E_{ij} ,
 \label{LLb=c}
\end{multline}
where $G_{i}=G^{(+)}_{i}+G^{(-)}_{i}$ is defined by 
\begin{align}
G^{(+)}_{i}&=q^{2e_{ii}}+(q-q^{-1})^{2}\sum_{k=1}^{i-1}e_{ki}\eb_{ik}q^{e_{kk}+e_{ii}-1},
\\[6pt]
G^{(-)}_{i}&=
q^{2(c-1)} \left(
q^{-2e_{ii}}+(q-q^{-1})^{2}
\sum_{k=i+1}^{N}e_{ki}\eb_{ik}q^{-e_{kk}-e_{ii}+1}
\right).
\end{align}
Here $A_{ji}$ and $C_{ji}$ for $i<j$ (resp.\ $i>j$) are the ones in 
\eqref{inter4a} (resp.\  \eqref{inter4b}) [dependence on the parameter $a$ is irrelevant].
 In deriving \eqref{LLb=c}, 
we have used \eqref{eeb1} and \eqref{eeb1-2}.
 Note that \eqref{LLb=c} becomes diagonal under the  relations 
\begin{align}
A_{ji}q^{2(c-1)}&=C_{ji}
\quad \text{for} \quad i  < j, \quad \text{and} \quad 
\quad
A_{ji}=C_{ji}q^{2(c-1)}
\quad \text{for} \quad j  < i.
\label{conAC}
\end{align}
We remark that  a part of the relations \eqref{con1} and \eqref{con2} automatically follows from \eqref{conAC}. 
For $N=2$, 
the relation \eqref{conAC} becomes trivial, and the diagonal elements 
$G_{1}$ and $G_{2}$ reduce to a Casimir element of $U_{q}(gl(2))$ 
[cf.\  eq.\ (3.5) in \cite{PT18}]. 
On the other hand,  for $N \ge 3$ case, 
each diagonal element is not a Casimir element, but rather 
 twisted sums
\footnote{$\sum_{k=1}^{N}G_{k}q^{2k}$ is 
a part of the  twisted trace of \eqref{LLb=c} 
(by $1 \otimes \pi(q^{2\sum_{k=1}^{N}ke_{kk}})$).} 
$\sum_{k=1}^{N}G^{(+)}_{k}q^{2k}$, 
$\sum_{k=1}^{N}G^{(-)}_{k}q^{2k}$ and $\sum_{k=1}^{N}G_{k}q^{2k}$ are 
(cf.\ eq.\ (41) in \cite{Zhang92}). 
 However, there are cases where simplifications occur for particular representations. 
 In fact, one can check that  the q-oscillator representation \eqref{HP-rep} 
 satisfies the condition \eqref{conAC}, and $G_{i}=q^{2m}+q^{-2}$ holds for 
 any $i$. This is one of the desirable properties (a sort of unitarity condition) 
 for construction of mutually commuting 
 transfer matrices.  
\section{Augmented q-Onsager algebra like symmetry}
\label{sec:q-AO}
 In this section, we will reconsider the intertwining relations in the previous section 
 and point out their connection to the generators of $U_{q}(\hat{gl}(N))$.  
 We will also mention our observation on an underlying symmetry behind them. 

We find that the intertwining relations \eqref{inter1b}, \eqref{inter2b}, \eqref{inter4a} and \eqref{inter4b} 
 can be written in terms of the generators of $U_{q}(\hat{gl}(N))$: 
\begin{align}
\mathsf{ev}_{x^{-1}}(Z_{ji}) \Kf(x)=\Kf(x) \overline{\mathsf{ev}}_{x}(Z_{ji}), 
 \label{intertwine}
\end{align}
where $Z_{ji}$ is defined as follows:
\\
For $1\le i=j \le N$, 
\begin{align}
Z_{ii}=k_{i}.
 \label{Z-0}
\end{align}
For $1\le j<i \le a $ or $a+1\le j<i \le N $ , 
\begin{align}
Z_{ji}=e_{[j,i-1]}.
\end{align}
For $1\le i<j \le a $ or $a+1\le i<j \le N $, 
\begin{align}
Z_{ji}=f_{[i,j-1]} .
\end{align}
For $ j \le a <i $,
\begin{align}
Z_{ji}&=\epsilon_{+}Z^{+}_{ji}+\epsilon_{-}Z^{-}_{ji},
\nonumber
\\[6pt]
Z^{+}_{ji}&=
\begin{cases}
[ [\overline{f}_{[1,j-1]},f_{N}]_{q^{-1}},f_{[i,N-1]}]_{q}q^{k_{j}-k_{i}+1}
&
 \text{for}  \quad 2 \le j \le a <i \le N-1
 \\[6pt]
 [ f_{N}, f_{[i,N-1]}]_{q}q^{k_{1}-k_{i}+1}
&
 \text{for}  \quad 1= j \le a <i \le N-1
  \\[6pt]
[\overline{f}_{[1,j-1]},f_{N}]_{q^{-1}}q^{k_{j}-k_{N}+1}
&
 \text{for}  \quad 2 \le  j \le a <i = N
   \\[6pt]
f_{N} q^{k_{1}-k_{N}+1}
&
 \text{for}  \quad 1 =  j \le a <i = N,
\end{cases}
 \label{Z+1}
\\[6pt]
Z^{-}_{ji}&=
\begin{cases}
[ \overline{e}_{[j,a]},e_{[a+1,i-1]}]_{q^{-1}} 
&
 \text{for}  \quad 1 \le j \le a <a+1<i \le N
 \\[6pt]
\overline{e}_{[j,a]}
&
 \text{for}  \quad 1 \le j \le a <a+1=i \le N
  \\[6pt]
e_{[a,i-1]}
&
 \text{for}  \quad 1 \le j = a <i \le N.
 \end{cases}
 \label{Z-1}
\end{align}
For $ i \le a <j $: 
\begin{align}
Z_{ji}&=\epsilon_{+}Z^{+}_{ji}+\epsilon_{-}Z^{-}_{ji},
\nonumber 
\\[6pt]
Z^{+}_{ji}&=
\begin{cases}
[ [\overline{e}_{[j,N-1]},[e_{N},e_{[1,i-1]}]_{q^{-1}}]_{q} 
&
 \text{for}  \quad 2 \le i \le a <j \le N-1
 \\[6pt]
 [\overline{e}_{[j,N-1]}, e_{N}]_{q}
&
 \text{for}  \quad 1= i \le a <j \le N-1
  \\[6pt]
[e_{N},e_{[1,i-1]}]_{q^{-1}}
&
 \text{for}  \quad 2 \le  i \le a <j = N
   \\[6pt]
e_{N}
&
 \text{for}  \quad 1 =  i \le a <j = N,
\end{cases}
 \label{Z+2}
\\[6pt]
Z^{-}_{ji}&=
\begin{cases}
[ \overline{f}_{[a+1,j-1]}, f_{[i,a]}]_{q^{-1}} q^{k_{j}-k_{i}+1}
&
 \text{for}  \quad 1 \le i \le a <a+1<j \le N
 \\[6pt]
f_{[i,a]}q^{k_{a+1}-k_{i}+1}
&
 \text{for}  \quad 1 \le i \le a <a+1=j \le N
  \\[6pt]
\overline{f}_{[a,j-1]}q^{k_{j}-k_{a}+1}
&
 \text{for}  \quad 1 \le i = a <j \le N.
 \end{cases}
 \label{Z-2}
\end{align}
Here we use the following notation for root vectors. For $i<j$, we define
\begin{align}
\begin{split}
e_{[i,j]}&=[e_{i},[e_{i+1}, \dots ,[e_{j-2},[e_{j-1},e_{j}]_{q^{-1}}]_{q^{-1}} \dots ]_{q^{-1}} ]_{q^{-1}},
\\[6pt]
\overline{e}_{[i,j]}&=[e_{i},[e_{i+1}, \dots ,[e_{j-2},[e_{j-1},e_{j}]_{q}]_{q} \dots ]_{q} ]_{q},
\\[6pt]
f_{[i,j]}&=[f_{j},[f_{j-1}, \dots ,[f_{i+2},[f_{i+1},f_{i}]_{q}]_{q} \dots ]_{q} ]_{q},
\\[6pt]
\overline{f}_{[i,j]}&=[f_{j},[f_{j-1}, \dots ,[f_{i+2},[f_{i+1},f_{i}]_{q^{-1}}]_{q^{-1}} \dots ]_{q^{-1}} ]_{q^{-1}},
\end{split}
\end{align}
and $e_{[i,i]}=\overline{e}_{[i,i]}=e_{i}$, $f_{[i,i]}=\overline{f}_{[i,i]}=f_{i}$. 
We remark that the third case in \eqref{Z-1} (resp.\ \eqref{Z-2}) 
 is a special case of the first or the second ones. 
 In deriving these, we have used \eqref{eeb1}-\eqref{brabra3}.

 We introduce generators composed of Cartan elements
 \begin{align}
 \kz^{+}_{ji}=\epsilon_{+}q^{\sum_{l=1}^{j}k_{l}-\sum_{l=i}^{N}k_{l}},
 \quad 
  \kz^{-}_{ji}=\epsilon_{-}q^{-\sum_{l=j}^{a}k_{l}+\sum_{l=a+1}^{i}k_{l}} 
  \quad 
  \text{for} \quad 1 \le j \le a <i \le N,
 \end{align}
and use notation
 $Z_{a,a+1}=\ez_{a}$, $Z_{N,1}=\ez_{N}$, $Z_{a+1,a}=\fz_{a}$, $Z_{1,N}=\fz_{N}$, 
 $\kz^{+}_{a}=\kz^{+}_{a,a+1}$,  $\kz^{-}_{a}=\kz^{-}_{a,a+1}$, 
 $\kz^{+}_{N}=\kz^{+}_{1,N}$,  $\kz^{-}_{N}=\kz^{-}_{1,N}$.
We have observed
\footnote{We have checked these by Mathematica.} the following commutation relations among generators 
\footnote{Note that the $q$-commutators $[A,[A,B]_{q_{1}}]_{q_{2}}$ 
 and $[A,[A,[A,B]_{q_{1}}]_{q_{2}}]_{q_{3}}$ 
are invariant under the permutation on $\{q_{1},q_{2}\}$  and  $\{q_{1},q_{2},q_{3}\}$, 
respectively.} 
for the q-oscillator representation   of $U_{q}(\hat{gl}(N))$ under the 
evaluation map \eqref{eva} and \eqref{HP-rep}, 
in addition to the relations \eqref{rel-afglN} restricted to the generators 
$\{e_{i},f_{i}\}_{i \ne a,N}$ and $\{k_{i} \}_{i=1}^{N}$. 
\begin{align}
& [\kz^{+}_{ji},\kz^{-}_{rs}]= [\kz^{\pm}_{ji},\kz^{\pm}_{rs}]= [\kz^{\pm}_{ji},k_{l}]=0 \quad \text{for} \ 1 \le j,r \le a <i ,s \le N, \quad 1 \le l \le N;
\\[10pt]
& [k_{l},Z_{ji}]=(\delta_{lj}-\delta_{il})Z_{ji} \quad \text{for} \quad 1 \le l,i,j \le N;
\\[10pt]
\begin{split}
& \kz^{+}_{ji}Z_{rs}= q^{\theta(1\le r \le j)-\theta(i\le r \le N)-\theta(1\le s \le j)+\theta(i\le s \le N)} Z_{rs} \kz^{+}_{ji},  
\\[6pt]
& \kz^{-}_{ji}Z_{rs}= q^{-\theta(j\le r \le a)+\theta(a+1\le r \le i)+\theta(j\le s \le a)-
\theta(a+1\le s \le i)} Z_{rs} \kz^{-}_{ji} \\
&\hspace{120pt} \text{for} \ 1 \le j \le a <i \le N, \quad 1 \le r,s \le N;
\end{split}
\label{aorel1}
\\[10pt]
\begin{split}
& [\ez_{a},[\ez_{a},[\ez_{a},\fz_{a}]_{q^2}]]_{q^{-2}}=
\tilde{\rho} \ez_{a}((\kz_{a}^{+})^{2}-(\kz_{a}^{-})^{2})\ez_{a},
\\[6pt]
& [\ez_{N},[\ez_{N},[\ez_{N},\fz_{N}]_{q^2}]]_{q^{-2}}=
\tilde{\rho} \ez_{N}((\kz_{N}^{-})^{2}-(\kz_{N}^{+})^{2})\ez_{N},
\\[6pt]
& [\fz_{a},[\fz_{a},[\fz_{a},\ez_{a}]_{q^2}]]_{q^{-2}}=
\tilde{\rho} \fz_{a}((\kz_{a}^{-})^{2}-(\kz_{a}^{+})^{2})\fz_{a},
\\[6pt]
& [\fz_{N},[\fz_{N},[\fz_{N},\ez_{N}]_{q^2}]]_{q^{-2}}=
\tilde{\rho} \fz_{N}((\kz_{N}^{+})^{2}-(\kz_{N}^{-})^{2})\fz_{N}
\quad \text{for} \quad 1 \le a \le N-1;
\end{split}
\label{aorel2}
\\[10pt]
\begin{split}
&
[\ez_{a},[\ez_{a},[\ez_{a},\ez_{N}]_{q^2}]]_{q^{-2}}=[\fz_{a},[\fz_{a},[\fz_{a},\fz_{N}]_{q^2}]]_{q^{-2}}=0,
\\[6pt]
&
[\ez_{N},[\ez_{N},[\ez_{N},\ez_{a}]_{q^2}]]_{q^{-2}}=
[\fz_{N},[\fz_{N},[\fz_{N},\fz_{a}]_{q^2}]]_{q^{-2}}=0 ,
\\[6pt]
& [\fz_{N},\ez_{a}]_{q^{-2}}=[\ez_{N},\fz_{a}]_{q^{2}}=0
\quad \text{for} \quad 2 \le a \le N-2;
\end{split}
\label{aorel3}
\\[10pt]
\begin{split}
&
[e_{a-1},[e_{a-1},\ez_{a}]_{q}]_{q^{-1}}=[\ez_{a},[\ez_{a},e_{a-1}]_{q}]_{q^{-1}}=
[f_{a-1},[f_{a-1},\fz_{a}]]_{q^{2}}=[\fz_{a},[\fz_{a},f_{a-1}]]_{q^{-2}}=0, 
\\[6pt]
&
[e_{1},[e_{1},\ez_{N}]_{q}]_{q^{-1}}=[\ez_{N},[\ez_{N},e_{1}]_{q}]_{q^{-1}}=
[f_{1},[f_{1},\fz_{N}]]_{q^{2}}=[\fz_{N},[\fz_{N},f_{1}]]_{q^{-2}}=0,
 \\[6pt]
& [e_{a-1},\fz_{a}]_{q^{-1}}= [e_{1},\fz_{N}]_{q^{-1}}=0 
\quad \text{for} \quad 2 \le a \le N-1;
\end{split}
\label{aorel4}
\\[10pt]
\begin{split}
&
[e_{a+1},[e_{a+1},\ez_{a}]_{q}]_{q^{-1}}=[\ez_{a},[\ez_{a},e_{a+1}]_{q}]_{q^{-1}}=
[f_{a+1},[f_{a+1},\fz_{a}]]_{q^{2}}=[\fz_{a},[\fz_{a},f_{a+1}]]_{q^{-2}}=0, 
\\[6pt]
&
[e_{N-1},[e_{N-1},\ez_{N}]_{q}]_{q^{-1}}=[\ez_{N},[\ez_{N},e_{N-1}]_{q}]_{q^{-1}}=0,
\\[6pt]
&
[f_{N-1},[f_{N-1},\fz_{N}]]_{q^{2}}=[\fz_{N},[\fz_{N},f_{N-1}]]_{q^{-2}}=0,
 \\[6pt]
&  [e_{a+1},\fz_{a}]_{q^{-1}}= [e_{N-1},\fz_{N}]_{q^{-1}}=0  
\quad \text{for} \quad 1 \le a \le N-2;
\end{split}
\label{aorel5}
\\[10pt]
&
[e_{i},\ez_{a}]=[e_{i},\fz_{a}]=[f_{i},\fz_{a}]=0  \quad 
\text{for} \quad 1 \le i \le a-2 \quad \text{or} \quad a+2 \le i \le N-1;
\label{aorel6}
\\[10pt]
&
[e_{i},\ez_{N}]=[e_{i},\fz_{N}]=[f_{i},\fz_{N}]=0  \quad 
\text{for} \quad 2 \le i \le a-1 \quad \text{or} \quad a+1 \le i \le N-2;
\label{aorel7}
\\[10pt]
&
[f_{i},\ez_{a}]=[f_{i},\ez_{N}]=0  \quad 
\text{for} \quad 1 \le i \le a-1 \quad \text{or} \quad a+1 \le i \le N-1;
\label{aorel8}
\\[10pt]
& [e_{i-1},Z_{ij}]_{q}=Z_{i-1,j}, 
\quad 
[Z_{ji},f_{i-1}]=Z_{j,i-1}q^{-1+k_{i-1}-k_{i}}
\quad \text{for} \quad 2 \le i \le a<j\le N;
\label{aorel9}
\\[10pt]
& [Z_{ij},e_{j}]_{q^{-1}}=Z_{i,j+1}, 
\quad 
[f_{j},Z_{ji}]=Z_{j+1,i}q^{k_{j}-k_{j+1}}
\quad \text{for} \quad 1 \le i \le a<j\le N-1;
\label{aorel10}
\\[10pt]
\begin{split}
& [\fz_{1},[\fz_{1},[\fz_{1},\fz_{N}]_{q^3}]_{q}]_{q^{-1}}=
q^{-1}\rho
 \fz_{1}(\kz_{1}^{-})^{2}\fz_{1}Z_{2N},
\\[6pt]
& [\fz_{N},[\fz_{N},[\fz_{N},\fz_{1}]_{q^{-3}}]_{q^{-1}}]_{q}=
-q^{-4}\rho
 \fz_{N}(\kz_{1}^{-})^{2}\fz_{N}Z_{2N} ,
\\[6pt]
& [\ez_{1},[\ez_{1},[\ez_{1},\ez_{N}]_{q^3}]_{q}]_{q^{-1}}=
-q^{5}\rho
 \ez_{1}(\kz_{N}^{+})^{2}\ez_{1}Z_{N2}q^{k_{N}-k_{2}},
\\[6pt]
& [\ez_{N},[\ez_{N},[\ez_{N},\ez_{1}]_{q^{-3}}]_{q^{-1}}]_{q}=
q^{2}\rho
 \ez_{N}(\kz_{N}^{+})^{2}\ez_{N}Z_{N2}q^{k_{N}-k_{2}} 
\quad \text{for}  \quad  a=1, \quad  N>2;
\end{split}
\label{LBBBa}
\\[10pt]
\begin{split}
& [ \fz_{N-1},[\fz_{N-1},[\fz_{N-1},\fz_{N}]_{q^3}]_{q}]_{q^{-1}}=
-q^{2}\rho
 \fz_{N-1}(\kz_{N}^{-})^{2} \fz_{N-1}Z_{1,N-1} q^{2k_{1}},
\\[6pt]
& [\fz_{N},[\fz_{N},[\fz_{N},\fz_{N-1}]_{q^{-3}}]_{q^{-1}}]_{q}=
q\rho
 \fz_{N}(\kz_{N}^{-})^{2}\fz_{N}Z_{1,N-1} q^{2k_{1}},
\\[6pt]
& [\ez_{N-1},[\ez_{N-1},[\ez_{N-1},\ez_{N}]_{q^3}]_{q}]_{q^{-1}}=
\rho
 \ez_{N-1}(\kz_{N-1}^{+})^{2}\ez_{N-1}Z_{N-1,1}q^{-k_{1}-k_{N-1}} ,
\\[6pt]
& [\ez_{N},[\ez_{N},[\ez_{N},\ez_{N-1}]_{q^{-3}}]_{q^{-1}}]_{q}=
-q^{-1}\rho
 \ez_{N}(\kz_{N-1}^{+})^{2}\ez_{N}Z_{N-1,1}q^{-k_{1}-k_{N-1}} 
 \\
 & \hspace{200pt}
\text{for} \quad  a=N-1,\quad  N>2,
\end{split}
\label{BBBaR}
\end{align}
where 
$\rho=(q^{3}-q^{-3})(q^{2}-q^{-2})$, 
$\tilde{\rho}=\rho /(q-q^{-1})$. 
We remark that the parameter $a$ is a fixed parameter in the diagonal K-matrix \eqref{K-dia}. 
The generators $\ez_{i}$ and $\fz_{i}$ are analogues
\footnote{This analogy is not perfect since  
they are mixtures of $e_{i}$ and $f_{i}$ as can be seen from 
\eqref{Z+1}-\eqref{Z-2}.} of 
the generators $e_{i}$ and $f_{i}$ of the quantum affine algebra. 
In fact, they satisfy Serre like relations (see, \eqref{aorel3}-\eqref{aorel5}). 
$Z_{ij}$ are analogue of higher root vectors, which are connected to the 
 generators $e_{i}$ and $f_{i}$ of the quantum affine algebra through \eqref{aorel10}. 
The relations \eqref{aorel2} are characteristic of this system, which look different 
from the relations of the quantum affine algebra; while their variants \eqref{LBBBa} 
and \eqref{BBBaR} exist only for $N>2$ case. 
These commutation relations are valid without any restriction of the class of 
representations for $N=2$ case, and define the so-called augmented q-Onsager algebra \cite{IT,BB1}, a realization of which   
in terms of the generators of $U_{q}(\hat{sl}(2))$  is known in \cite{BB1}.  
In contrast, not all of them hold true for generic representations for $N \ge 3$ case. 
Whether the elements $Z_{ji}$ for $N \ge 3$ 
 satisfy closed commutation relations on the level of 
 the algebra 
 remains to be clarified.
\section{Rational limit}
 In this section, we consider the rational limit of some of the formulas in previous sections.

Let $\epsilon_{\pm}=\mp q^{-2p_{\pm}}$, $p=p_{+}-p_{-}$, and $u \in \mathbb{C}$. 
In order to take the 
 rational limit of the  L-operators \eqref{Lop} and \eqref{Lbop},  
 we  must renormalize them by the factor 
 $(q-q^{-1})^{-1}$. The factor is necessary to get finite non-zero results:
\begin{align}
\Ls(u) &=\lim_{q \to 1} (q-q^{-1})^{-1}\Lf(q^{-2u}) =\sum_{i,j}
(su \delta_{ij}+e_{ji}) \otimes E_{ij},
\\[6pt]
\Lbs(u) &=\lim_{q \to 1} (q-q^{-1})^{-1}\Lbf(q^{-2u}) =\sum_{i,j}
(-su \delta_{ij}+e_{ji}) \otimes E_{ij},
\end{align}
where $e_{ij}$ are generators of $gl(N)$:
\begin{align}
[e_{ij},e_{kl}]=\delta_{jk}e_{il}-\delta_{li}e_{kj}. 
  \label{rel-glN}
\end{align}
 The limit of the K-matrix \eqref{K-dia} can be taken similarly:
\begin{align}
{\mathit K}(u) &=\lim_{q \to 1} (q-q^{-1})^{-1}K(q^{-2u}) =\sum_{k=1}^{a}
(su +p)E_{kk}+
\sum_{k=a+1}^{N}(-su+p)E_{kk}.
\end{align}
As for the K-operators \eqref{K-op}-\eqref{K-op4}, 
we rewrite them in terms of a q-analogue of the gamma function 
(see for example, \cite{AAR99})  
\begin{align}
\Gamma_{q}(x)=
\frac{(q;q)_{\infty}}{(q^{x};q)_{\infty}} (1-q)^{1-x} 
\qquad \text{for} 
\qquad |q| <1,
 \label{q-gamma}
\end{align}
which reduces to the normal gamma function in the rational limit:
$\lim_{q \to 1} \Gamma_{q}(x)=\Gamma (x)$. 
The limit $\Ks(u)=\lim_{|q| \to 1 + 0}\Kf(q^{-2u})(1-q^{- 2})^{1-c+2su}$ 
for \eqref{K-op} and the limit 
$\Ks(u)=\lim_{|q| \to 1 - 0}\Kf(q^{-2u})(1-q^{2})^{1-c+2su}$  for  \eqref{K-op4} 
give the same result:
\begin{align}
\Ks(u)&=\frac{\Gamma \left(-su-p+\sum_{k=1}^{a}e_{kk}\right) }
{\Gamma \left(su-p+1-\sum_{k=a+1}^{N}e_{kk}\right) } .
  \label{Kop-ra}
\end{align}
The limit $\Ks(u)=\lim_{|q| \to 1 - 0}\Kf(q^{-2u})(1-q^{2})^{1-c+2su}$
for \eqref{K-op2} and the limit 
 $\Ks(u)=\lim_{|q| \to 1 + 0}\Kf(q^{-2u})(1-q^{- 2})^{1-c+2su}$  for  \eqref{K-op3} 
 give the same result:
\begin{align}
\Ks(u)& =\frac{\Gamma \left(-su+p+\sum_{k=a+1}^{N}e_{kk}\right) }
{\Gamma \left(su+p+1-\sum_{k=1}^{a}e_{kk}\right) }.
  \label{Kop-ra2}
\end{align}
Then, after renormalizing the K-operator as above, the limit of 
the intertwining relations \eqref{inter1b} and \eqref{inter2b} 
can be taken straightforwardly:
\begin{align}
& e_{ji}\Ks(u) =\Ks(u)e_{ji}
\quad  \text{for} \quad  i,j \le a \quad \text{or} \quad i,j \ge a+1.
 \label{rel-ra1}
\end{align}
As for the  intertwining relations  \eqref{inter4a} and \eqref{inter4b}, 
 in addition to renormalizing the K-operator, 
one has to divide both sides of them by $(q-q^{-1})$ before taking the limit 
to get:
\begin{align}
& \left( (su-p)e_{ji} -\sum_{k=a+1}^{N}e_{ki} e_{jk} \right)  \Ks(u) =
\Ks(u) \left( (-su-p)e_{ji} +\sum_{k=1}^{a}e_{ki} e_{jk} \right)  
\nonumber \\
& \hspace{200pt}  \text{for} \quad  i \le a<j ,
\\[6pt]
& \left( (-su-p)e_{ji} +\sum_{k=1}^{a}e_{ki} e_{jk} \right)  \Ks(u) =
\Ks(u) \left( (su-p)e_{ji} -\sum_{k=a+1}^{N}e_{ki} e_{jk} \right)  
\nonumber \\
&  \hspace{200pt}  \text{for} \quad  j \le a<i .
\label{rel-ra3}
\end{align}
We find that  \eqref{Kop-ra} and \eqref{Kop-ra2}  
solve the relations \eqref{rel-ra1}-\eqref{rel-ra3}, and thus the reflection equation 
\begin{multline}
\Ls_{12} \left(v-u \right) \Ks_{1}(u) \overline{\Ls}_{12} \left( u+v \right) 
 {\mathit K}_{2}(v) 
={\mathit K}_{2}(v) 
 \Ls_{12} \left(-u-v \right)  \Ks_{1}(u) \overline{\Ls}_{12} \left(u-v\right) 
 \\
 \text{for} \quad u,v \in \mathbb{C},
\label{refeq2rat}
\end{multline}
which is the rational limit of \eqref{refeq2}, if the following conditions
\footnote{These are sufficient conditions. If the parameter $p$ (in addition to $u$) is interpreted as 
a free one, these are necessary conditions as well. 
On the other hand, \eqref{con1} and \eqref{con2} are necessary and sufficient conditions independent of $\epsilon_{\pm}$.}
 are satisfied.
\begin{align}
e_{ji} \sum_{k=a+1}^{N}e_{kk}=\sum_{k=a+1}^{N}e_{ki}e_{jk}, 
\quad 
e_{ji}\left( \sum_{k=1}^{a}e_{kk}-1\right)=\sum_{k=1}^{a}e_{ki}e_{jk} 
\quad \text{for} \quad i \le a < j, 
 \label{sufcon-ra1}
\\[6pt]
e_{ji} \sum_{k=1}^{a}e_{kk}=\sum_{k=1}^{a}e_{ki}e_{jk}, 
\quad 
e_{ji}\left( \sum_{k=a+1}^{N}e_{kk}-1\right)=\sum_{k=a+1}^{N}e_{ki}e_{jk} 
\quad \text{for} \quad j \le a < i.
 \label{sufcon-ra2}
\end{align}
The relations \eqref{sufcon-ra1} and  \eqref{sufcon-ra2} correspond 
  to the rational limit of
 \eqref{con1} and \eqref{con2}, respectively. 
 Then the rational limit of \eqref{HP-rep} satisfies \eqref{sufcon-ra1} and \eqref{sufcon-ra2}.
Note that the above relations reproduce (a part of) a kind of generalized rectangular condition
\footnote{In case the representations are finite dimensional, this condition 
 is satisfied by the representations labeled by rectangular Young diagrams.}
\cite{FLMS13}. 
\begin{align}
\sum_{k=1}^{N}e_{ki}e_{jk}=\alpha e_{ji} +\beta \delta_{ij} 
\qquad \alpha, \beta \in {\mathbb C}.
 \label{rek}
\end{align}
In fact, sum of the first and the second relations in \eqref{sufcon-ra1} or 
\eqref{sufcon-ra2} gives
\begin{align}
e_{ji}\left( \sum_{k=1}^{N}e_{kk}-1\right)=\sum_{k=1}^{N}e_{ki}e_{jk} 
\qquad \text{for} \quad  i \le a < j \quad \text{or} \quad  j \le a < i .
\end{align}
This corresponds to \eqref{rek} for $\beta =0$ since 
 $\alpha:=\sum_{k=1}^{N}e_{kk}-1$ is a central element of $gl(N)$.
\section{Concluding remarks}
We have derived intertwining relations \eqref{inter1b}, \eqref{inter2b}, \eqref{inter4a} 
and \eqref{inter4b} from the reflection equation \eqref{refeq2} for L-operators  
associated with  $U_{q}(\hat{gl}(N))$ 
and have obtained the diagonal K-operators \eqref{K-op}-\eqref{K-op4} 
in terms of the Cartan elements of a quotient of $U_{q}(gl(N))$ 
 by the relations \eqref{con1} and \eqref{con2}. 
The intertwining relations can be expressed in terms of generators of $U_{q}(\hat{gl}(N))$ 
as in \eqref{intertwine}. 
In particular, the elements $\{ Z_{ji} \}$ 
 \eqref{Z-0}-\eqref{Z-2} in the intertwining relations  
 satisfy augmented q-Onsager algebra like commutation relations for the 
 q-oscillator representation \eqref{HP-rep} under the evaluation map \eqref{eva}.
 It will be an interesting problem to investigate 
 the algebra generated by  $\{ Z_{ji} \}$ (with some constraint on generators of 
 $U_{q}(\hat{gl}(N))$, if any), 
 which could be a version of higher rank generalization of the augmented q-Onsager algebra. 
 The reflection equation \eqref{refeq0} 
 defines \cite{Skly} the reflection equation algebra
 if one adds the third component in the K-matrix:
\begin{align}
\begin{split}
& R_{12} \left(\frac{y}{x}\right) K_{13}(x) \overline{R}_{12} \left( xy \right) 
 K_{23}(y) 
=K_{23}(y) 
 R_{12} \left(\frac{1}{xy} \right)  K_{13}(x) \overline{R}_{12} \left(\frac{x}{y}\right), 
\\[6pt]
 &K_{13}(x)=\sum_{i,j=1}^{N}E_{ij} \otimes 1 \otimes k_{ij}(x), 
 \qquad 
  K_{23}(x)=\sum_{i,j=1}^{N}1  \otimes E_{ij} \otimes k_{ij}(x),
  \end{split}
  \label{refeq03}
 \end{align}
 where the elements $ \{k_{ij}(x)\}$ generate the algebra. 
 In this context, 
 it will be desirable to clarify the direct connection 
 between $\{ Z_{ji} \}$ and a certain specialization of $\{ k_{ij}(x)\}$ for the diagonal 
 K-operators. 

In this paper, we have defined the L-operator $\Lfb(x)$ based on the second evaluation 
map \eqref{eva2}. 
It will be worthwhile to define it based on the first one \eqref{eva} 
as $\Lfb^{\prime} (xy^{-1})=\overline{\phi}^{\prime}(xy^{-1})(\mathsf{ev}_{x} \otimes \pi_{y}) {\mathcal R}_{21}$ and investigate solutions of the reflection equation.  
The reason why we avoided $\Lfb^{\prime} (x)$ was that its expression based on the 
generators $e_{ij}$ or $\eb_{ij}$ might be involved for $N \ge 3$ case. The conditions 
\eqref{con1} and \eqref{con2} for the solutions of the reflection equation
 are the ones that  the L-operator $\Lfb(x)$ coincides with $\Lfb^{\prime}(x)$ 
up to a fine tune by the central element (see the relation \eqref{ev-evb}). 
The technical difficulty on $\Lfb^{\prime} (x)$ may be resolved if we start 
from the FRT formulation of the algebra (based on Yang-Baxter relations) \cite{Faddeev:1987ih} 
 and use the matrix elements of the 
L-operators as the generators of the algebra instead of $e_{ij}$ or $\eb_{ij}$. 

Another important problem is construction of Baxter Q-operators for 
open boundary conditions in the light of \cite{FS15,PT18}. 
The K-operators obtained in this paper could be building blocks 
of Q-operators after taking limits on the parameter $m$ of the q-oscillator representation, 
as was already discussed in \cite{PT18} for $N=2$ case. 

\section*{Acknowledgments} 
The author would like to thank Pascal Baseilhac, Sergey Khoroshkin and 
Masato Okado  for correspondence. 
He also thanks the anonymous referees for comments. 


\section*{Appendix A: Relations for $U_{q}(gl(N))$}
\label{relationsglmn}
\addcontentsline{toc}{section}{Appendix A}
\def\theequation{A\arabic{equation}}
\setcounter{equation}{0}
Here we review relations among the generators of $U_{q}(gl(N))$, 
some of which are used in the main text. 
\begin{align}
\begin{split}
& e_{ab}=[e_{ac},e_{cb}]_{q} \quad \text{for} \quad a>c>b,
\\
& e_{ab}=[e_{ac},e_{cb}]_{q^{-1}} \quad \text{for} \quad a<c<b,
\\
& [e_{ab}, e_{ba}] = 
 \frac{q^{ e_{aa} - e_{bb} } -q^{- e_{aa} + e_{bb} }  }{q-q^{-1}}
 \quad \text{for} \quad a <b,
\\
& [e_{dc}, e_{ba}] =- (q-q^{-1}) e_{da} e_{bc} 
\quad \text{for} \quad b <d<a<c 
\\
& \hspace{130pt} \text{or} \quad 
a<c<b<d ,
\\
& [e_{dc}, e_{ba}] =0
\quad \text{for} 
\quad d <c<b<a \quad \text{or}  
\quad d >c>b>a \quad \text{or} 
\quad d <b<a<c \quad \text{or} 
\\
& 
\quad d >b>a>c \quad \text{or} 
\quad d <c \le a<b\quad \text{or} 
\quad c <d \le b<a\quad \text{or} 
\quad d <a < b<c \quad \text{or} 
\\
& 
\quad c <b < a<d ,
\\
& [e_{dc}, e_{ba}] =- (q-q^{-1}) 
 q^{ e_{aa} - e_{cc} } e_{da} e_{bc} 
\\
& 
 \quad \text{for} \quad d <a<c<b,   
\\
& [e_{dc}, e_{ba}] = (q-q^{-1}) 
e_{da} e_{bc}  q^{ e_{bb} - e_{dd} } 
\\
& 
 \quad \text{for} \quad a <d<b<c,     
\\
& [e_{ba}, e_{ac}] = e_{bc}  q^{ e_{bb} - e_{aa} } 
 \quad \text{for} \quad a <b<c,  
\\
& [e_{ba}, e_{ac}] =  q^{ e_{aa} - e_{cc} } e_{bc} 
 \quad \text{for} \quad a <c<b,    
\\
& [e_{db}, e_{ba}] = e_{da}  q^{e_{bb} - e_{dd} } 
 \quad \text{for} \quad a <d<b,  
\\
& [e_{db}, e_{ba}] = q^{e_{aa} -e_{bb} } e_{da}  
 \quad \text{for} \quad d <a<b,  
\\
& [e_{da}, e_{ba}] _{q^{-1}}=0 
 \quad \text{for} \quad a <b<d
  \quad \text{or} \quad b <d<a,
\\
& [e_{bc}, e_{ba}] _{q}=0 
 \quad \text{for} \quad c <a<b
  \quad \text{or} \quad b <c<a.
  \end{split}
\end{align}
The relations for $\eb_{ij}$ can be obtained from the above relations 
through $q \to q^{-1}$. In addition, one can prove the following relations 
based on the above relations and  induction. 
We remark that  \eqref{eeb1-2} for $l=j-1$ corresponds to eq.\  (10) in \cite{Zhang92}.  
\begin{align}
e_{ji}-(q-q^{-1})\sum_{k=j+1}^{l}e_{ki}\eb_{jk}&=
\eb_{ji}+(q-q^{-1})\sum_{k=l+1}^{i-1}e_{ki}\eb_{jk}=
\nonumber 
\\ 
&=
\begin{cases}
 \eb_{ji} & \text{for} \quad j\le l=i-1, \\
 [\eb_{j,l+1},e_{l+1,i}]_{q^{-1}} & \text{for} \quad j \le l < i-1, \\
 e_{ji} &  \text{for} \quad j=l<i,
\end{cases}
 \label{eeb1}
\\[6pt]
\eb_{ji}-(q-q^{-1})\sum_{k=i+1}^{l}e_{ki}\eb_{jk}&=
e_{ji}+(q-q^{-1})\sum_{k=l+1}^{j-1}e_{ki}\eb_{jk}=
\nonumber 
\\ 
&=
\begin{cases}
 e_{ji} & \text{for} \quad i\le l=j-1, \\
 [\eb_{j,l+1},e_{l+1,i}]_{q^{-1}} & \text{for} \quad i \le l < j-1, \\
 \eb_{ji} &  \text{for} \quad i=l<j,
\end{cases}
 \label{eeb1-2}
\\[6pt]
e_{ji}+(q-q^{-1})\sum_{k=l}^{j-1}e_{ki}\eb_{jk}q^{e_{kk}-e_{jj}-1}&=
 [\eb_{jl},e_{li}]_{q^{-2}}q^{e_{ll}-e_{jj}} \quad \text{for} \quad l < j < i, 
 \label{eeb2}
\\[6pt]
\eb_{ji}+(q-q^{-1})\sum_{k=l}^{i-1}e_{ki}\eb_{jk}q^{e_{kk}-e_{ii}}&=
 [\eb_{jl},e_{li}]_{q^{-2}}q^{e_{ll}-e_{ii}+1} \quad \text{for} \quad l < i < j, 
 \label{eeb2-2}
\\[6pt]
\eb_{ji}-(q-q^{-1})\sum_{k=i+1}^{l}e_{ki}\eb_{jk}q^{e_{ii}-e_{kk}}&=
 [\eb_{jl},e_{li}]_{q^{2}}q^{e_{ii}-e_{ll}-1} \quad \text{for} \quad j < i < l, 
 \label{eeb3}
\\[6pt]
e_{ji}-(q-q^{-1})\sum_{k=j+1}^{l}e_{ki}\eb_{jk}q^{e_{jj}-e_{kk}+1}&=
 [\eb_{jl},e_{li}]_{q^{2}}q^{e_{jj}-e_{ll}} \quad  \text{for} \quad i < j < l, 
 \label{eeb3-2}
\\[6pt]
  [\eb_{jl},e_{li}]_{q^{2}}= [\eb_{j,l+1},e_{l+1,i}]q^{e_{ll}-e_{l+1,l+1}} &
\quad \text{for} \quad i<j<l \quad \text{or} \quad j<i<l, 
\label{brabra1}
\\[6pt]
  [\eb_{jl},e_{li}]_{q^{-2}}= [\eb_{j,l-1},e_{l-1,i}]q^{e_{l-1,l-1}-e_{ll}} &
\quad \text{for} \quad l<i<j \quad \text{or} \quad l<j<i, 
\label{brabra2}
\\[6pt]
  [\eb_{j,l-1},e_{l-1,i}]_{q}= [\eb_{jl},e_{li}]_{q^{-1}} \quad  
\text{for}  \quad  i&<l-1<l<j  \quad \text{or} \quad j<l-1<l<i.
\label{brabra3}
\end{align}
We also remark that the third relation in the right hand side of
 \eqref{eeb1} (and  \eqref{eeb1-2}) is a special case of the second one.

\end{document}